\documentclass[journal]{IEEEtran}

\usepackage[utf8]{inputenc}

\usepackage{amsmath,amssymb}

\usepackage[dvipsnames]{xcolor}
\usepackage{graphicx}
\usepackage{tikz}
\usepackage{pgfplots}
\usepackage{caption}
\DeclareCaptionFont{eightpt}{\fontsize{8}{9.6}\selectfont}
\usepackage{float}
\usepackage{balance}
\usetikzlibrary{patterns}
\usetikzlibrary{arrows.meta}
\usetikzlibrary{arrows}
\usepackage{diagbox}
\usepackage{hyperref}
\hypersetup{colorlinks,linkcolor=red,citecolor=blue} 

\usepackage{xspace}
\usepackage[normalem]{ulem}
\usepackage{dsfont,empheq}
\usepackage{multicol,multirow}

\usepackage{cite}
\usepackage{relsize}

\usetikzlibrary{plotmarks,matrix,chains,scopes,fit,calc,shapes,positioning,decorations,intersections,arrows,backgrounds,shadows}

\newlength\FigureHeight
\newlength\FigureWidth
\pgfkeys{/pgf/number format/.cd,1000 sep={}}
\pgfplotsset{every axis legend/.append style={legend cell align=left,at={(0.02,0.97)},anchor=north west}}
\pgfplotsset{every axis plot/.append style={line width=1.5pt}} 
\pgfplotsset{every axis plot/.append style={mark options={solid,fill=white!80!.,line width=0.5pt},mark size = 3pt}} 

\definecolor{myDarkGreen}{rgb}{0.00000,0.58824,0.00000}%
\definecolor{uniform}{rgb}{0.00000,0.58824,0.00000}%
\definecolor{matched}{rgb}{1,0.58824,0.00000}%
\definecolor{mismatched}{rgb}{0.00000,0.44700,0.74100}%
\definecolor{AWGNreference}{rgb}{0.00000,0.58824,0.00000}%
\definecolor{XPMmodel}{rgb}{0,0,0}%
\definecolor{sims}{rgb}{0.00000,0.44700,0.74100}%

\pgfplotsset{AWGN_capacity/.style={color=gray,dotted}}
\pgfplotsset{16QAM_uniform/.style={color=blue,dashed}}
\pgfplotsset{16QAM_shaped/.style={color=blue,solid}}
\pgfplotsset{64QAM_uniform/.style={color=red,dashed}}
\pgfplotsset{64QAM_shaped/.style={color=red,solid}}
\pgfplotsset{256QAM_uniform/.style={color=brown,dashed}}
\pgfplotsset{256QAM_shaped/.style={color=brown,solid}}

\pgfplotsset{AWGNreference/.style={color=AWGNreference,dotted}}
\pgfplotsset{XPMmodel/.style={color=XPMmodel,dashed}}
\pgfplotsset{sims/.style={color=sims,only marks,mark=*,mark options={solid,fill=white}}}

\newcounter{lemma}

\newtheorem{exampleplain}{Example}

\newenvironment{example}{\begin{exampleplain}}{~\hfill$\vartriangle$\end{exampleplain}}

\pgfplotsset{compat=newest} 

\begin{document}
\title{Improved Decoding of Staircase Codes: The Soft-aided Bit-marking (SABM) Algorithm}

\markboth{Preprint, \today.}{}

\author{Yi Lei, Bin Chen,~\IEEEmembership{Member,~IEEE}, Gabriele Liga, Xiong Deng, Zizeng Cao, Jianqiang Li,~\IEEEmembership{Senior Member,~IEEE}, Kun Xu,~\IEEEmembership{Member,~IEEE}, Alex Alvarado,~\IEEEmembership{Senior Member,~IEEE}
\thanks{Y. Lei is with State Key Laboratory of Information of Photonics and Optical Communications, Beijing University of Posts and Telecommunications (BUPT), China and the Signal Processing Systems (SPS) Group, Department of Electrical Engineering, Eindhoven University of Technology, The Netherlands (\mbox{e-mail:}leiyi@bupt.edu.cn).}
\thanks{A. Alvarado, B. Chen, G. Liga and X. Deng are with the Signal Processing Systems (SPS) Group, Department of Electrical Engineering, Eindhoven University of Technology, The Netherlands. Z. Cao and B. Chen are with the Electro-Optical Communications (ECO) Group, Department of Electrical Engineering, Eindhoven University of Technology, The Netherlands (\mbox{e-mails:} \{a.alvarado, b.c.chen,g.liga,x.deng,z.cao\}@tue.nl).}
\thanks{J. Li and K. Xu are with State Key Laboratory of Information of Photonics and Optical Communications, Beijing University of Posts and Telecommunications (BUPT), China (\mbox{e-mails: }jianqiangli@bupt.edu.cn, xukun@bupt.edu.cn)}

\thanks{This work was presented in part at the International Symposium on Turbo Codes \& Iterative Information Processing 2018, Hong Kong, China, Dec. 2018.}
}

\maketitle

\begin{abstract}
Staircase codes (SCCs) are typically decoded using iterative bounded-distance decoding (BDD) and hard decisions. In this paper, a novel decoding algorithm is proposed, which partially uses soft information from the channel. The proposed algorithm is based on marking certain number of highly reliable and highly unreliable bits. These marked bits are used to improve the miscorrection-detection capability of the SCC decoder and the error-correcting capability of BDD. For SCCs with $2$-error-correcting Bose-Chaudhuri-Hocquenghem component codes, our algorithm improves upon standard SCC decoding by up to $0.30$~dB at a bit-error rate (BER) of $10^{-7}$. The proposed algorithm is shown to achieve almost half of the gain achievable by an idealized decoder with this structure. 
A complexity analysis based on the number of additional calls to the component BDD decoder shows that the relative complexity increase is only around $4\%$ at a BER of $10^{-4}$. This additional complexity is shown to decrease as the channel quality improves. Our algorithm is also extended (with minor modifications) to product codes. The simulation results show that in this case, the algorithm offers gains of up to $0.44$~dB at a BER of $10^{-8}$.
\end{abstract}


\section{Introduction and Motivation}
Forward error correction (FEC) is required in optical communication systems to meet the ever increasing data demands in optical transport networks (OTNs). FEC codes that can boost the net coding gain (NCG) are of key importance. A Reed-Solomon (RS) code with parameters of $(255,239)$ was the first standardized FEC code for OTNs in the ITU-T Recommendation G.975 \cite{G975}. For an output bit error ratio (BER) of $10^{-15}$, the NCG of RS$(255,239)$ is $6.2$~dB. In order to increase transmission data rate and distance, several super FEC codes were considered in the ITU-T Recommendation G.975.1 \cite{G9751}. Most of these super FECs utilize two concatenated FEC codes, such as: BCH$(3860,3824,3)$+BCH$(2040,1930,10)$ codes\footnote{Throughout this paper we use $n_{c}$, $k_{c}$, and $t$ to denote the codeword length, information length, and error-correcting capability, resp. BCH codes are denoted as BCH$(n_{c},k_{c},t)$.}
, RS$(1023,1007)$+BCH$(2047,1952,8)$ codes, etc. The achieved NCG can be up to $8.99$~dB at a BER of $10^{-15}$.

OTNs are currently targeting data rates of $400$~Gb/s and beyond \cite{400G1,400G3}. In this scenario, FEC codes with  higher NCG are highly desired. Soft-decision (SD) FEC codes provide large NCGs, however, they are not the best candidates for very high data rate applications due to their high power consumption and decoding latency. 
For applications with strict latency and complexity requirements (e.g., short reach), simple but powerful hard-decision (HD) FEC codes, e.g., product codes (PCs)\cite{JornTComm2011} and staircase codes (SCCs) \cite{Smith2012,Zhang2014}, have received considerable attention: PC has been adopted (as an inner code) in the subclass I.$5$ of G.975.1 \cite{G9751}, while SCC is part of the $400$ZR Implementation Agreement (as an outer code) in the Optical Internetworking Forum \cite{OIF400G}. SCC is also recommended for $100$G optical transport unit (OTU) order $4$ for long-reach applications in the ITU-T Recommendation G.709.2/Y.1331.2 \cite{G709}. 
In \cite{FougstedtJLt2019}, product and staircase decodes are  implemented in very-large-scale integration system, which reach more than 1~Tb/s information throughputs with only energy efficiencies of around 2~pJ/bit.
Recently, low-complexity concatenated FEC and adaptive coded modulation schemes have also been studied to combine the advantages of soft- and hard-decision decoders \cite{BarakatainJLT2018,BinACP2018}.

Both SCCs and PCs are based on simple component codes, Bose-Chaudhuri-Hocquenghem (BCH) codes being the most popular ones. The decoding is done iteratively based on bounded-distance decoding (BDD) for the component codes. Although very simple, one drawback of BDD is that its error-correcting capability is limited to $t=\lfloor(d_{0}-1)/2\rfloor$, where $d_{0}$ is the minimum Hamming distance (MHD) of the component code\cite{BDD}. 
BDD can detect more than $t$ errors, but cannot correct them. In some cases, BDD may also erroneously decode a received sequence with more than $t$ errors, a situation known as a \emph{miscorrection}. Miscorrections are known to degrade the performance of iterative BDD. To prevent miscorrections and/or extend the error correcting capability, several methods have been studied in the literature. In what follows we review those methods.

To prevent miscorrections in SCCs, the authors of \cite{SmithPhD} proposed rejecting bit-flips from the decoding of bit sequences associated with the last SCC block if they conflict with a zero-syndrome codeword from the previous block. As pointed out in \cite[Sec.~I]{ChristianJournal}, the obtained gains of \cite{SmithPhD} are expected to be limited. An anchor-based decoding algorithm has been proposed in \cite{ChristianJournal,Christian1}, where some bit sequences are labeled as anchor codewords. These sequences are thought to have been decoded without miscorrections. Decoding results that are inconsistent with anchor codewords are discarded. 
It has been demonstrated that this algorithm works well with both SCCs and PCs. The algorithm of \cite{Christian1} outperforms \cite{SmithPhD}, but it suffers from an increased complexity as anchor codewords need to be tracked during iterative BDD. Very recently, a modified iterative BDD for PCs was proposed in \cite{Alireza,AlirezaISTC,AlirezaOFC2019,AlirezaarXiv2019}. In this algorithm, channel reliabilities are used to perform the final HD at the output of BDD, instead of directly accepting the decoding result. Large gains are obtained, but it requires additional memory (and processing) as all the soft information needs to be saved. Moreover, its effectiveness for SCCs has not yet been reported in the literature.

To extend the error correcting capability, Chase proposed three kinds of algorithms to decode block codes with channel soft information \cite{ChaseDecoding}. In this class of algorithm, each bit is accompanied with an analog weight, according to the soft information. During the decoding, the algorithms will generate a sequence of test patterns first, then decode all of them and choose the decoding result with lowest analog weight as the final output. Through this way, the error correcting capability can be extended from $\lfloor(d_{0}-1)/2\rfloor$ to $d_{0}-1$. The main drawback of these three algorithms is that the decoder needs to decode at least $\lfloor(d_{0}/2)+1\rfloor$ test patterns (while in fact, not all of them are necessary). This significantly increases the decoding complexity and latency. In addition, algorithm $3$ in \cite{ChaseDecoding} behaves similarly to erasure decoding that the sequence of test patterns is equivalent to the sequence of erasures described in \cite[Sec.~6.6]{ErasureDecoding}. Both Chase decoders and erasure decoding were not designed to take miscorrections into account. In addition, based on chase decoder, the authors of \cite{Douxin_ISTC2018} have considered to use soft-input/soft-output decoder to decode each component code within all iterations. However, the achieved additional gain is only $0.30$~dB for large block size, at the expense of greatly increased complexity.

In this paper, we propose the soft-aided bit-marking (SABM) algorithm to improve the decoding of SCCs as well as PCs. As high order modulation formats are often used in modern optical communication systems, the performance of the proposed SABM algorithm under different modulation formats was investigated. The presented gains are achieved by marking highly reliable and highly unreliable bits, an idea we first proposed in \cite{YiISTC2018} and also experimentally validated in a multi-span hybrid-amplified system in \cite{BinOFC2019}. Unlike previous works, the proposed SABM algorithm \emph{jointly} increases the miscorrection-detection capability of the SCC decoder and the error-correcting capability of BDD. The main feature of the proposed algorithm is its low complexity, as explained in what follows. For SCCs, the SABM algorithm only requires modifications to the decoding structure of the last block of each decoding window. Furthermore, in the SABM algorithm each component code needs to be decoded at most twice. Also, the algorithm is based on marking bits only, and thus, no soft bits (log-likelihood ratios, LLRs) need to be stored. Finally, marked bits do not need to be tracked during the iterative process either. 

The remainder of the paper is organized as follows. In Sec. II, we present the system model we consider and introduce the principles of SCCs and BDD. In Sec. III, we describe the proposed SABM algorithm. Some examples are also gave to explain how it works. In Sec. IV, we present the simulation results for SCCs, and discuss the complexity of the SABM algorithm. In Sec. V, we extend this algorithm to PCs and investigate the performance. Finally, we conclude this paper in Sec. VI.

\section{System Model, SCCs, and BDD}\label{Sec:General}

\subsection{System Model}

\begin{figure}[t]
\centering
\includegraphics[width=0.48\textwidth]{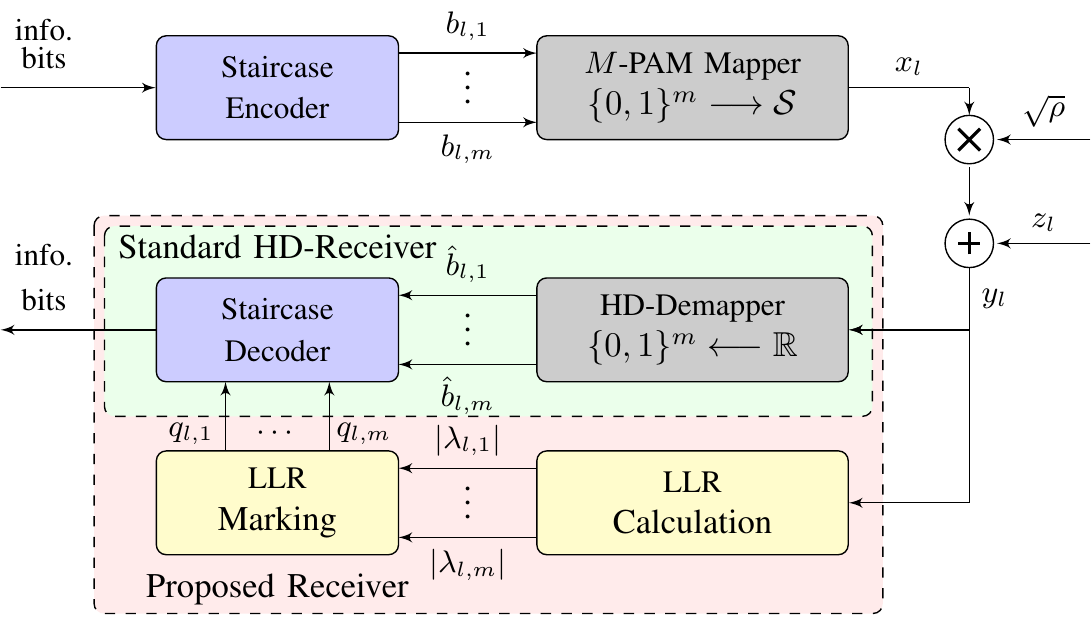}
\caption{System model under consideration.}
\label{fig: System Model}
\end{figure}

As shown in Fig. \ref{fig: System Model}, information bits are encoded by a staircase encoder and then mapped to symbols $x_{l}$ taken from an equally-spaced $M$-ary Pulse Amplitude Modulation (PAM) constellation $\mathcal{S}=\{s_{1},s_{2},\ldots,s_{M}\}$ with $M=2^m$ points, where $l$ is the discrete time index. The bit-to-symbol mapping is the binary reflected Gray code. The received signal is ${y_{l}}=\sqrt{\rho}{x_{l}}+{z_{l}}$, where ${z_{l}}$ is zero-mean unit-variance additive white Gaussian noise (AWGN). 

The standard HD receiver structure for SCCs uses an HD-based demapper to estimate the code bits, which are then fed to the decoder (green area in Fig.~\ref{fig: System Model}). In this paper, we consider a receiver architecture where the HD-FEC decoder uses soft information from the channel. This soft information is typically represented using LLRs, calculated as \cite[eq.~(3.50)]{LLR}
\begin{equation}\label{LLR}
   \lambda_{l,k}=\sum_{b \in \{0,1\}} (-1)^{\bar{b}} \log\sum_{i \in \mathcal{I}_{k,b}} \textrm{exp}\left(-\frac{(y_{l}-\sqrt{\rho}s_{i})^{2}}{2}\right),
\end{equation}
with $k=1,\ldots,m$, and where $\bar{b}$ denotes bit negation. In \eqref{LLR}, the set $\mathcal{I}_{k,b}$ enumerates all the constellation points in $\mathcal{S}$ whose $k$th bit $c_{i,k}$ is $b$, i.e., $\mathcal{I}_{k,b}\triangleq \{i=1,2,\ldots,M: c_{i,k}=b\}$.

The proposed structure is shown in Fig.~\ref{fig: System Model} (red area). In this structure, apart from the HD-estimated bits $\hat{b}_{l,1},\ldots,\hat{b}_{l,m}$, a sequence of \emph{marked} bits will also be made available to the decoder. We call this architecture soft-aided (SA) HD-FEC decoding. These marked bits are denoted by $q_{l,k}$ and can be: highly reliable bits (HRBs), highly unreliable bits (HUBs), or unmarked bits. The marking is made based on the absolute value of the LLRs $|\lambda_{l,k}|$. More details about the marking procedure and how this can be exploited by the decoder will be given in Sec.~\ref{sec:algorithm}.

\subsection{Staircase Codes}

\begin{figure}[tpb]
\centering
\includegraphics[width=0.35\textwidth]{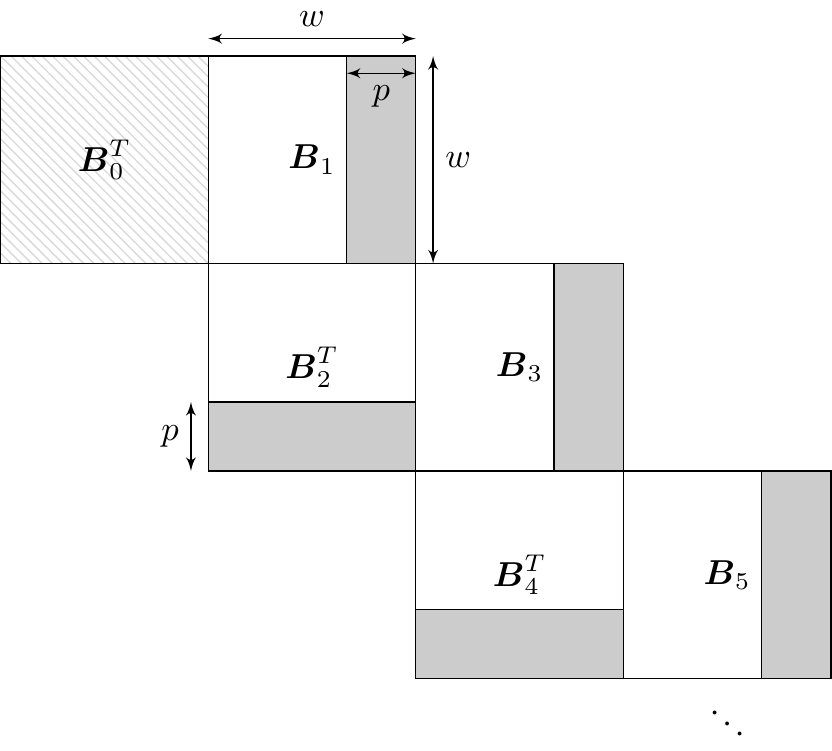}
    \caption{Staircase structure of SCCs considered in this paper.}
   \label{fig: Structure of SCC}
\end{figure}

Fig. \ref{fig: Structure of SCC} shows the staircase structure of SCCs we consider in this paper, where block $\boldsymbol{B}_{0}$ is initialized to all zeros. Each subsequent SCC block $\boldsymbol{B}_{i},i=1,2,\ldots$, is composed of $w(w-p)$ information bits (white areas) and $wp$ parity bits (gray areas). Each row of the matrix $[\boldsymbol{B}^{T}_{i-1} \boldsymbol{B}_{i}]$ $\forall i>1$ is a valid codeword in a component code $\mathcal{C}$. We consider the component code $\mathcal{C}$ to be a binary code with parameters $(n_{c}, k_{c}, t)$. 
Then, $w$ and $p$ are given by: $w=n_{c}/2$ and $p=n_{c}-k_{c}$. The code rate $R$ of the SCC is $R=1-p/w=2k_{c}/n_{c}-1$. Throughout this paper, the component codes $\mathcal{C}$ considered are extended (by 1 additional parity bit) BCH codes. The mapping between code bits and symbols is done by reading row-by-row the SCC blocks $\boldsymbol{B}_{i},i=1,2,\ldots$

At the receiver side, SCCs are decoded iteratively using a sliding window covering $L$ blocks. We use $\boldsymbol{Y}_{i}$ to indicate the received SCC block after HD-demapper corresponding to the transmitted block $\boldsymbol{B}_{i}$. The decoder first iteratively decodes the blocks $\{\boldsymbol{Y}_{0},\boldsymbol{Y}_{1},\ldots, \boldsymbol{Y}_{L-1}\}$. When a maximum number of iterations $\ell$ is reached, the decoding window outputs the block $\boldsymbol{Y}_{0}$ and moves to decode the blocks $\{\boldsymbol{Y}_{1}, \boldsymbol{Y}_{2},\ldots, \boldsymbol{Y}_{L}\}$. The block $\boldsymbol{Y}_{1}$ is then delivered and operation continues on $\{\boldsymbol{Y}_{2}, \boldsymbol{Y}_{3},\ldots, \boldsymbol{Y}_{L+1}\}$. This process continues indefinitely. Multiple decoding scheduling alternatives exist (see, e.g., \cite[Sec.~IV]{Smith2012}\cite[Sec.~II]{Zhang2014}). We chose the most popular one, namely, alternated decoding of pairs of SCC blocks within a window, from the bottom right to the top left of the SCC window.

\subsection{Bounded-Distance Decoding}

BDD is used to decode (in Hamming space) the received bit sequence for the component code $\mathcal{C}$. To correct up to $t$ errors, the MHD $d_{0}$ of $\mathcal{C}$ must satisfy $d_{0}\geq 2t+1$ ($d_{0}\geq 2t+2$ for extended BCH codes with 1 additional parity bit). Thus, every codeword in the code $\mathcal{C}$ can be associated to a sphere of radius $t$. Within such a sphere, no other codewords exist. If the received sequence $r$ falls inside one of these spheres, BDD will decode $r$ to the corresponding codeword. Otherwise, BDD will declare a failure. For a given transmitted codeword $c$ and a received sequence $r$, the BDD output $\hat{c}$ is thus given by
\begin{equation}\label{BDDequation}
     \begin{aligned}
     \hat{c}&= \left\{
     \begin{array}{lcl}
     c,  &      & \textrm{if~} d_{\textrm{H}}(r,c) \leq t  \\
    \tilde{c} \in \mathcal{C},  &      & \textrm{if~} d_{\textrm{H}}(r,c) > t \textrm{~and~} d_{\textrm{H}}(r,\tilde{c}) \leq t\\
     r, &      & d_{\textrm{H}}(r,\tilde{c}) > t~\forall \tilde{c} \in \mathcal{C} \\
     \end{array}
     \right.
     \end{aligned},
\end{equation}
 where $d_{\textrm{H}}(\cdot,\cdot)$ represents the Hamming distance. In practice, BDD is often a syndrome-based decoder that uses syndromes to estimate the error pattern $e$. If the syndromes are all zeros, no errors are present. For the first two cases in \eqref{BDDequation}, BDD will both declare decoding success and $\hat{c}=r\oplus e$. In the second case, although BDD will still return an error pattern $e$, this case corresponds to a miscorrection. In the next section, we will show how to improve miscorrection detection (MD) using the underlying structure of SCCs and the marked HRBs.

\section{The SABM Algorithm}\label{sec:algorithm}

\begin{figure}[tb]
 \centering
\includegraphics[width=.48\textwidth]{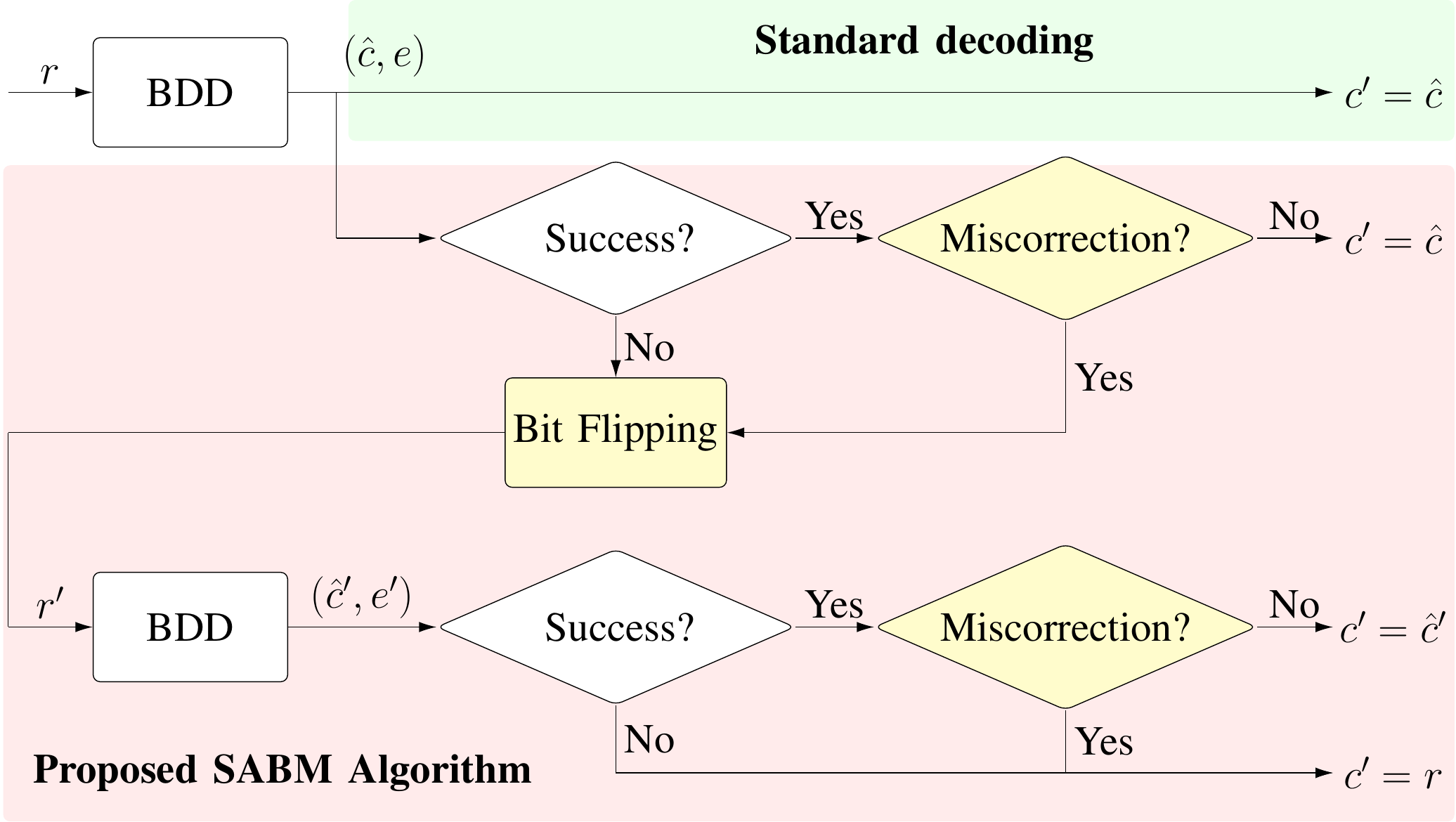}
\caption{Schematic diagram of the proposed SABM algorithm: $r$ is the received row sequence taken from two neighbor SCC blocks and $c'$ is the output of the staircase decoder. BDD returns a decoded codeword $\hat{c}$ based on \eqref{BDDequation} and an error pattern $e$. The three highlighted blocks use marked bits for operation.}
 \label{fig: Method}
\end{figure}

The schematic diagram of the proposed SABM algorithm is shown in Fig. \ref{fig: Method} (red area). Assume that decoding is being performed over the blocks $\{\boldsymbol{Y}_{i}, \boldsymbol{Y}_{i+1}, ... ,\boldsymbol{Y}_{i+L-1}\}$, then $r$ is given by a row sequence taken from two neighbor blocks $[\boldsymbol{Y}^{T}_{i+s-1} \boldsymbol{Y}_{i+s}]$, where $s\in\{1,2,\ldots,L-1\}$. Compared to standard SCC decoding (green area in Fig. \ref{fig: Method}), which always accepts the decoding result $\hat{c}$ of BDD, the SABM algorithm further checks the decoding status of BDD. If BDD successfully decodes $r$, miscorrection detection is performed. Furthermore, bit flipping (BF) is proposed as a way to handle decoding failures and miscorrections. In this section, we will explain the steps in the SABM algorithm.  

The SABM algorithm can in principle be applied to all received sequences $r$ within $L$ SCC blocks. However, due to the iterative sliding window decoding structure applied to SCCs, most of the errors are expected to be located in the last two blocks. To keep the complexity and latency low, we will therefore only use this algorithm on the received sequences from the last two blocks of the window. Therefore, from now on we only consider rows of the matrix $[\boldsymbol{Y}^{T}_{i+L-2} \boldsymbol{Y}_{i+L-1}]$.

\subsection{Decoding Success: Improved Miscorrection Detection}

To avoid miscorrections, it was suggested in \cite{SmithPhD} to reject the decoding result of BDD applied to $[\boldsymbol{Y}^{T}_{i+L-2} \boldsymbol{Y}_{i+L-1}]$ if the decoded codeword would cause conflicts with zero-syndrome codewords in $[\boldsymbol{Y}^{T}_{i+L-3} \boldsymbol{Y}_{i+L-2}]$. This method protects bits in $\boldsymbol{Y}_{i+L-2}$ but cannot handle bits in the last block $\boldsymbol{Y}_{i+L-1}$. We propose to enhance this method by using marked bits in $\boldsymbol{Y}_{i+L-1}$. In particular, we add one additional constraint to the algorithm in \cite{SmithPhD}: no HRBs in $\boldsymbol{Y}_{i+L-1}$ shall ever be flipped.

The reliability of a bit is given by the absolute value of its LLR, a high value indicating a more reliable bit. Therefore, a threshold $\delta$ is set to decide if the bit is highly reliable. If $|\lambda_{l,k}| > \delta$, the corresponding bit is marked as an HRB. The decision of the staircase decoder will therefore be marked as a miscorrection if the decoded codeword causes conflicts with zero-syndrome codewords in $[\boldsymbol{Y}^{T}_{i+L-3} \boldsymbol{Y}_{i+L-2}]$, \emph{or} if the decoded codeword flips a bit whose LLR satisfies $|\lambda_{l,k}| > \delta$. 

\begin{figure}[tbp]
\centering
\includegraphics[width=0.4\textwidth]{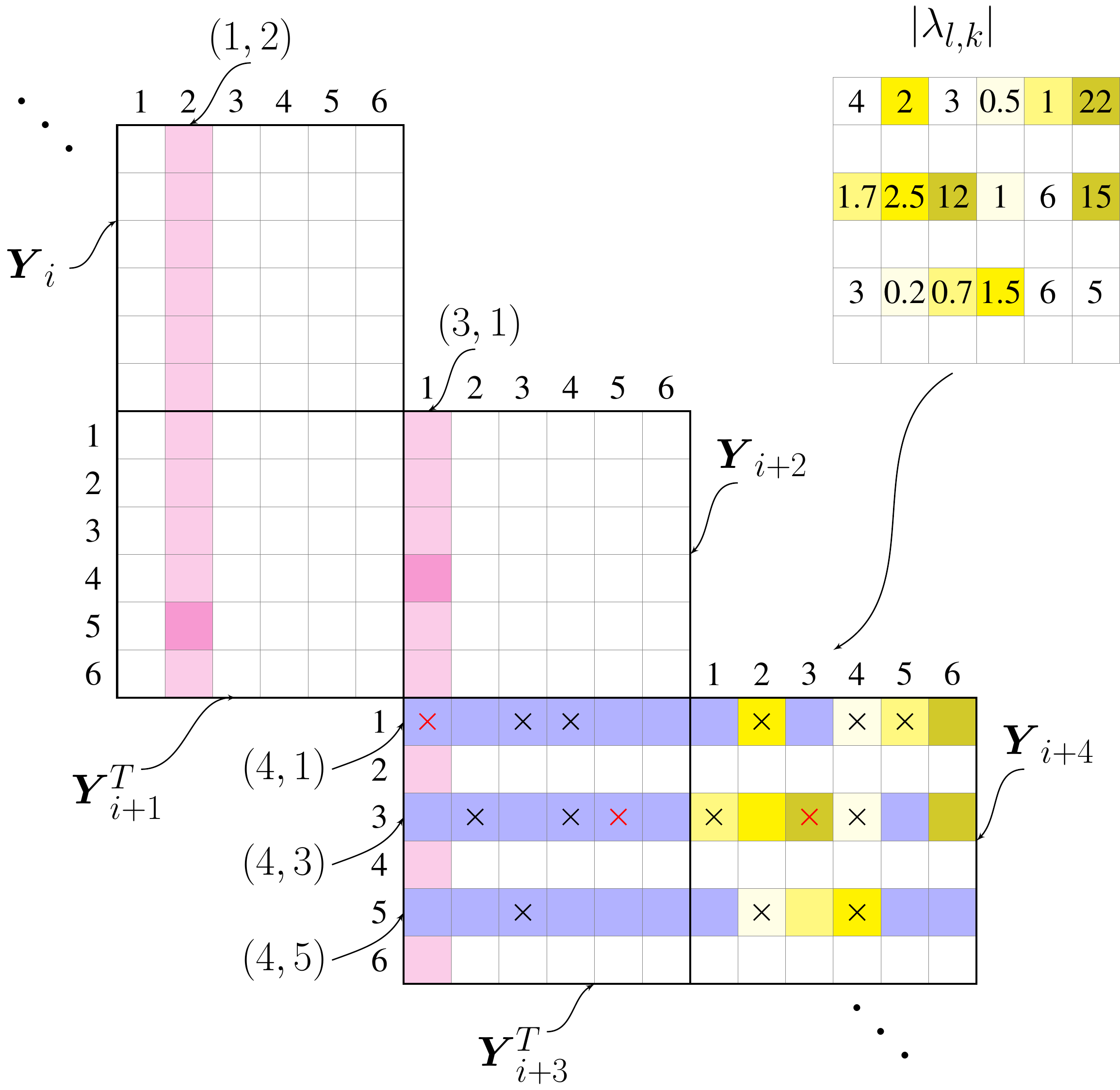}
 \caption {Decoding example ($w=6$, $L=5$, $t=2$): black crosses are received errors after channel transmission and red crosses indicate miscorrections after BDD. Dark yellow cells are marked HRBs ($\delta=10$), while light yellow cells are marked HUBs. Lighter yellow colors indicate a smaller value of $|\lambda_{l,k}|$.}
 \label{fig: SCC-example}
 \end{figure}
 
\begin{example}\label{Example.1}
Fig. \ref{fig: SCC-example} shows a decoding window with $w=6$ and $L=5$ and a component code $\mathcal{C}$ with $t=2$ ($d_{0}=6$). Following the notation of \cite{Christian1}, a pair $(i,j)$ is used to specify the location of a component codeword in each window, where $i\in \{1,2,\ldots,L-1\}$ indicates the position relative to the current window and $j \in \{1,2,\ldots, w\}$ indicates the corresponding row or column index in the matrix of two neighbor blocks. A triple $(i,j,k)$ is used to indicate the $k$th bit in the component codeword $(i,j)$, where $k \in \{1,2,\ldots,2w\}$. For example, the component codewords $(1,2)$ and $(3,1)$ are highlighted with light magenta, while bits $(1,2,11)$ and $(3,1,4)$ are highlighted with dark magenta. The bit sequence $(3,1)$ is a codeword in $[\boldsymbol{Y}^{T}_{i+2} \boldsymbol{Y}_{i+3}]$ whose syndrome is equal to zero. The cells filled with dark yellow are the ones marked as HRBs whose $|\lambda_{l,k}|$ is more than $\delta$, while $\delta=10$. 

After transmission, the received bit sequences for $(4,1)$ and $(4,3)$ have $5$ and $4$ errors (black crosses), respectively. When applying BDD, miscorrections (red crosses) occur. For the received bits in $(4,1)$, BDD mistakenly detects bit $(4,1,1)$ as an error and suggests to flip it. However, because it is involved in the zero-syndrome codeword $(3,1),$ it will be identified as a miscorrection by both our MD algorithm and by the one in \cite{SmithPhD}. For the received bits in $(4,3)$, however, the suggested flipping bit $(4,3,5)$ in $\boldsymbol{Y}_{i+L-2}$ is not involved in any zero-syndrome codewords, and thus, \cite{SmithPhD} would fail to detect this miscorrection. The bit $(4,3,9)$ is a HRB, and thus, our MD algorithm will successfully identify it as a miscorrection.
\end{example}

The MD algorithm in \cite{SmithPhD} does not always detect the miscorrections. The new rule we introduced (never flip HRBs in $\boldsymbol{Y}_{i+L-1}$) is only heuristic and does not guarantee perfect MD either. For example, our MD algorithm fails when no bits are flipped by BDD because $r=\tilde{c}\in\mathcal{C}$. Nevertheless, as we will see later, our MD algorithm combined with bit flipping (see next Sec.) gives remarkably good results with very small added complexity.

\subsection{Decoding Failures and Miscorrections: Bit Flipping}

To deal with decoding failures and miscorrections, we propose to flip bits (see BF block in Fig.~\ref{fig: Method}). The main idea is to flip certain bits in $r$ and make the resulting sequence $r'$ (after BF) closer to $c$ in Hamming space. In particular, the proposed BF aims at making the Hamming distance between $r'$ and $c$ equal to $t$, so that BDD can correct $r'$ to the transmitted codeword $c$. Two cases are considered by our proposed algorithm: (1) decoding failures, and (2) miscorrections.

\textbf{Case 1 (Decoding Failures):} We target received sequences with $t+1$ errors. In this case, we flip a HUB with the lowest absolute LLR. The intuition here is that this marked bit was indeed one flipped by the channel. In the cases where the marked HUB corresponds to a channel error, the error correction capability of the code $\mathcal{C}$ is effectively increased by $1$ bit.

\textbf{Case 2 (Miscorrections):} We target miscorrections where BDD chooses a codeword $\tilde{c}\in\mathcal{C}$ at MHD of $c$. The intuition here is that most of the miscorrections caused by BDD will result in codewords at MHD from the transmitted codeword. When a miscorrection has been detected, our algorithm calculates the number of errors detected by BDD. This is equal to $d_{\textrm{H}}(r,\tilde{c})= w_{\textrm{H}}(e)$. Then, our algorithm flips $d_{0}-w_{\textrm{H}}(e)-t$ bits, which in \emph{some cases} will result in $r'$ that satisfy $d_{\textrm{H}}(c,r')=t$. This will lead BDD to find the correct codeword. More details are given in Examples~\ref{Example.2} and~\ref{Example.3}.
Again using the intuition that bits with the lowest reliability are the most likely channel errors, our BF algorithm flips the most unreliable $d_{0}-w_{\textrm{H}}(e)-t$ bits. In practice, this means that out of $n_c$ code bits per codeword, only $d_{0}-w_{\textrm{H}}(e)-t < t+1$ (or $t+2$ for extended BCH codes) HUBs need to be marked (and sorted). The BF block (see Fig.~\ref{fig: Method}) chooses the number of marked bits to flip based on this sorted list and the Hamming weight of the error pattern. 

\begin{figure}[tb]
\centering
\includegraphics[width=0.4\textwidth]{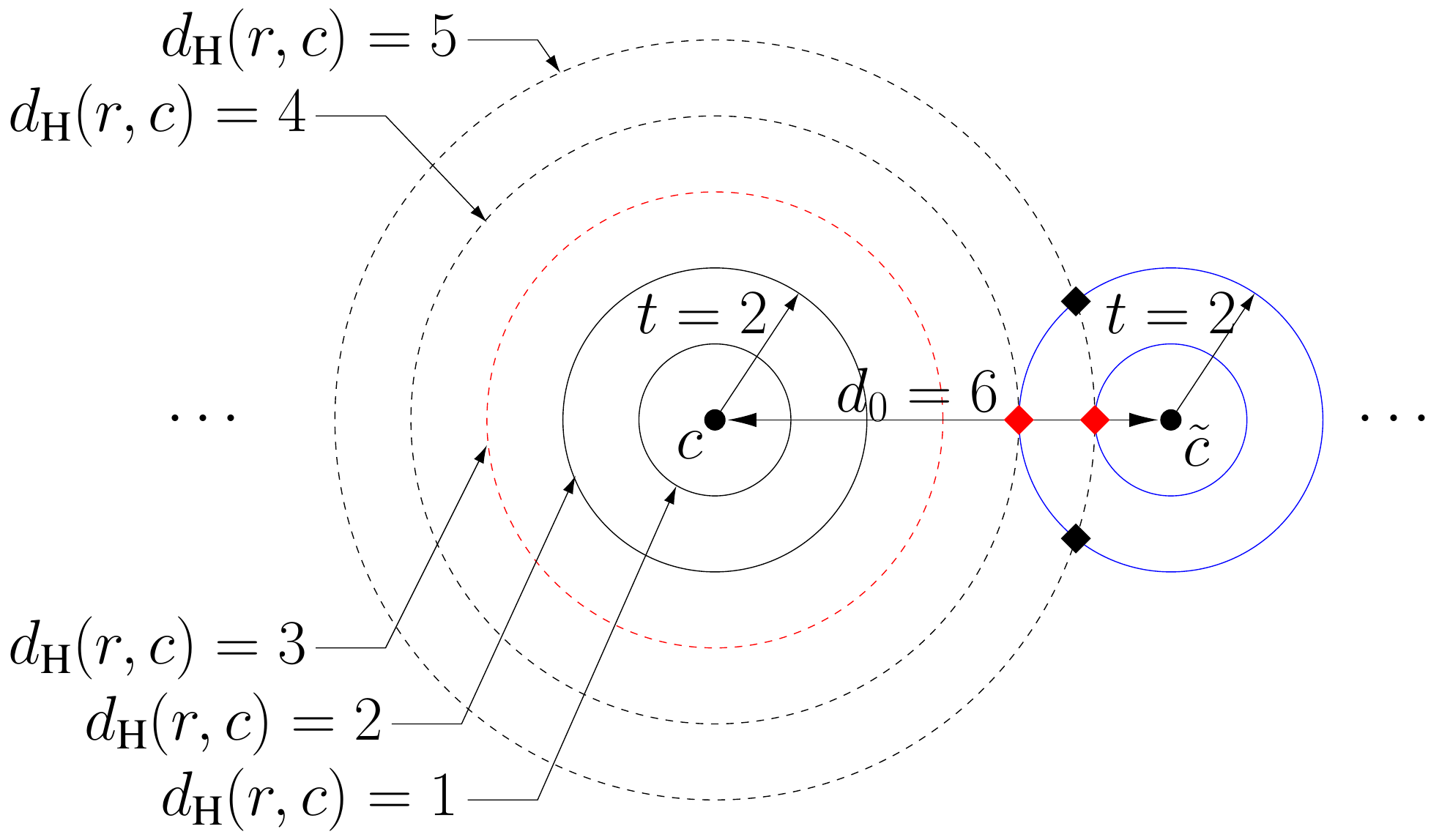}
\caption {Schematic representation of BDD: $c$ is the transmitted codeword and $\tilde{c}\in\mathcal{C}$ is another codeword at MHD $d_{0}=6$. The circles around $c$ show the possible locations of $r$ with 1, 2, .... errors from inside to outside in turn, while black solid circles indicate cases that BDD will decode successfully. Diamonds indicate four possible locations, where miscorrection happens.
}
\label{fig: BDD-example}
\end{figure}

\begin{example}\label{Example.2}
Fig. \ref{fig: BDD-example} shows a representation of BDD ($t=2$ and $d_{0}=6$), where the black dots represent the transmitted codeword $c$ and another codeword $\tilde{c}\in\mathcal{C}$ with $d_{\textrm{H}}(c,\tilde{c})=d_{0}$. The red dashed circle and solid blue circles correspond to locations of $r$ for Cases 1 and 2, respectively. The bit sequence $(4,5)$ in Fig.~\ref{fig: SCC-example} ($3$ errors) would lie on the red dashed circle, while sequences $(4,1)$ and $(4,3)$ correspond to red diamonds ($5$ and $4$ errors, respectively). For the latter two bit sequences, provided that we flip the correct bits (flipping $3$ and $2$ marked bits, respectively), will give a $r'$ with $d_{\textrm{H}}(c,r')=t$ which can be correctly decoded. 
\end{example}

\begin{example}\label{Example.3}
Light yellow cells in Fig. \ref{fig: SCC-example} show the marked 3 HUBs with the lowest reliability within that codeword. The lighter yellow color indicates a smaller value of $|\lambda_{l,k}|$. In this example, BDD fails to decode bit sequence $(4,5)$. Fortunately, $(4,5,8)$ corresponds to the marked HUB with smallest $|\lambda_{l,k}|$. Thus, it will be flipped after BF, and then the remaining 2 errors $(4,5,3)$ and $(4,5,10)$ will be fully corrected by applying BDD again. This corresponds to Case 1.

 For bit sequences (4,1) and (4,3), the decoding results of BDD are identified as miscorrections (as explained in Example \ref{Example.1}) with $w_{\textrm{H}}(e)=1$ and $w_{\textrm{H}}(e)=2$, respectively. According to the BF rule for miscorrections, 3 and 2 bits with smallest $|\lambda_{l,k}|$ among the marked HUBs, i.e., (4,1,8), (4,1,10), (4,1,11) in (4,1), and (4,3,7), (4,3,10) in (4,3), will all be flipped. As a result, only 2 errors are left in (4,1) and (4,3), which are within the error correcting capability of BDD. This corresponds to Case 2.
\end{example}

BF will not always result in the correct decision. As shown in Example~\ref{Example.2}, this is the case for certain miscorrections (black diamonds in Fig.~\ref{fig: SCC-example}). Additionally, miscorrections for codewords at distances larger than $d_0$ are not considered either. Finally, marked LLRs might not correspond to channel errors. In all these cases, either decoding failures or miscorrections will happen. To avoid these cases, the SABM algorithm includes two final checks after BF and BDD (see lowest part of Fig.~\ref{fig: Method}): successful decoding and MD.

\section{Algorithm Optimization and Simulation Results}

In this section, the component codes used for simulations are extended BCH codes with $1$ extra parity bit and 2-error-correcting capability ($t=2$). The decoding window size is $L=9$, and the maximum number of iterations is $\ell=7$. 

\subsection{LLR Threshold Choice}\label{sec:LLRchoice}

One key aspect of the proposed SABM algorithm is the selection of the bits to be marked as HRBs. This selection is based on the channel reliabilities, in particular, by using an LLR threshold $\delta$. In order to optimize the process of marking bits as highly reliable, the optimum LLR threshold need to be investigated. We do this in the following.

\begin{figure}[tpb]
\centering
\includegraphics[width=0.48\textwidth]{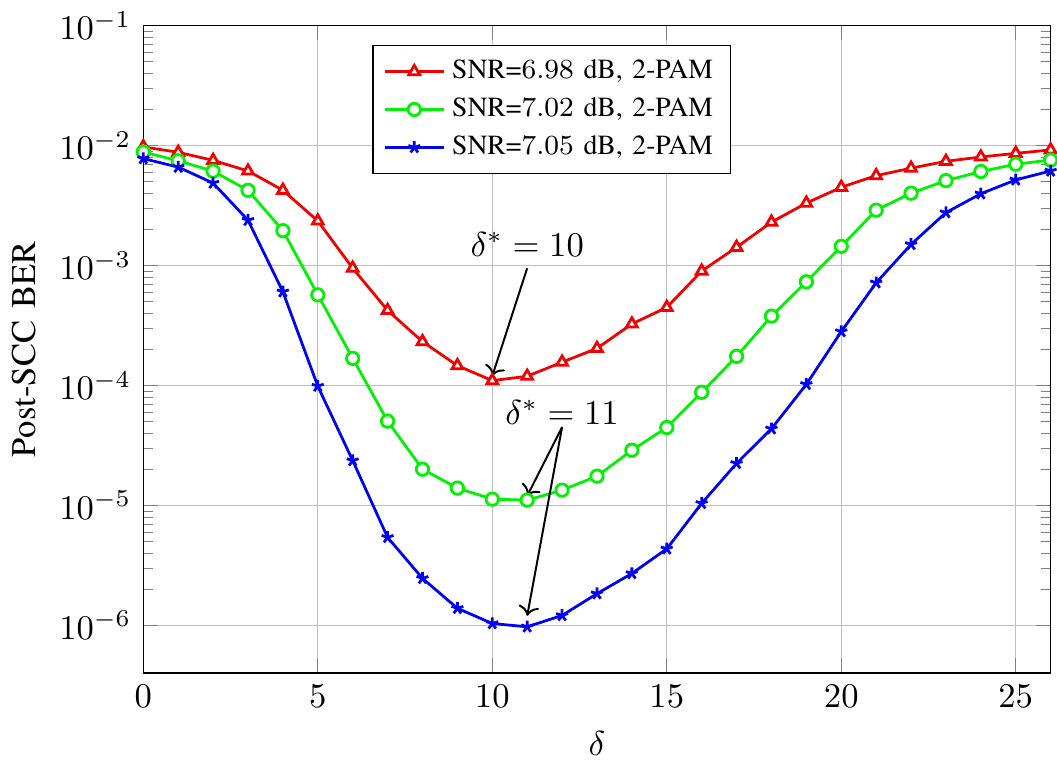}
\caption{Post-SCC BER vs. LLR threshold $\delta$ for code rate $R=0.87$ and $2$-PAM.}
\label{fig: Delta for R=0.87}
\end{figure}

To directly compare our algorithm with the results presented in \cite{Christian1}, we firstly consider an SCC with $R=0.87$, whose component code is BCH$(256,239,2)$ ($w=128$). Fig. \ref{fig: Delta for R=0.87} shows the post BER performance under different threshold $\delta$. The modulation format is 2-PAM. The three curves are obtained for SNRs of $6.98$~dB (triangles), $7.02$~dB (circles) and $7.05$~dB (stars), respectively. These SNRs are chosen so that the achieved BERs are $10^{-4}$, $10^{-5}$ and $10^{-6}$, resp. Fig. \ref{fig: Delta for R=0.87} shows that, to obtain the best performance, the corresponding optimum threshold $\delta^*$ is $10$, $11$ and $11$. However, the difference between these values is small, and the resulting performance difference is negligible as long as $\delta \approx 10$.\footnote{It is important to note that this difference becomes important for optical transmission experiments. This was recently shown in \cite[Fig.~3]{BinOFC2019}, where the optimum value $\delta^*$ for a long-haul system was found to be as low as $\delta^*=4$.}

The U-type trend results in Fig. \ref{fig: Delta for R=0.87} can be intuitively understood as follows. If $\delta < \delta^*$ is used, the performance degrades because some of the bits that are not reliable enough are marked as HRBs. This will lead to some correct BDD decisions being mistakenly marked as miscorrections, which are then rejected by the SABM-based staircase decoder. On the other hand, the performance degradation for $\delta > \delta^*$ is due to the fact that some of the bits that should probably be trusted, are not marked as HRBs. This weakens the ability of the SABM algorithm to identify miscorrections. 

\begin{figure}[tpb]
\centering
\includegraphics[width=0.48\textwidth]{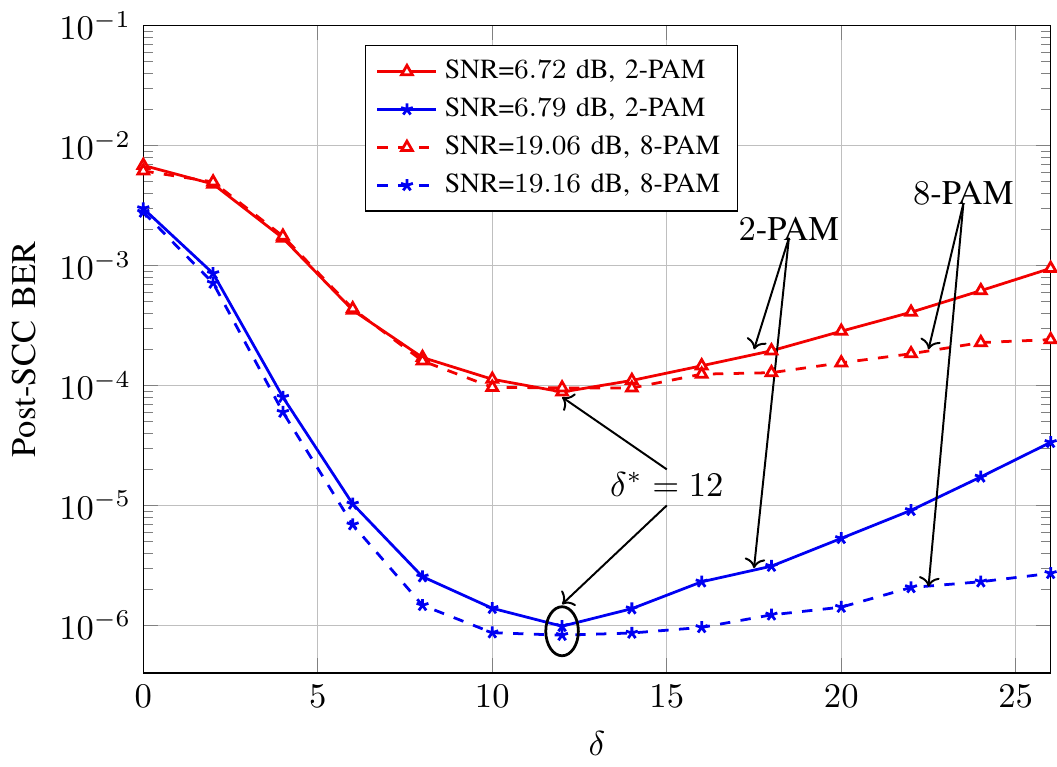}
\caption{Post-SCC BER vs. LLR threshold $\delta$ for code rate $R=0.83$. The modulation formats include $2$-PAM (solid lines) and $8$-PAM (dashed lines).}
\label{fig: Delta for R=0.83}
\end{figure}

\begin{figure}[tpb]
\centering
\includegraphics[width=0.48\textwidth]{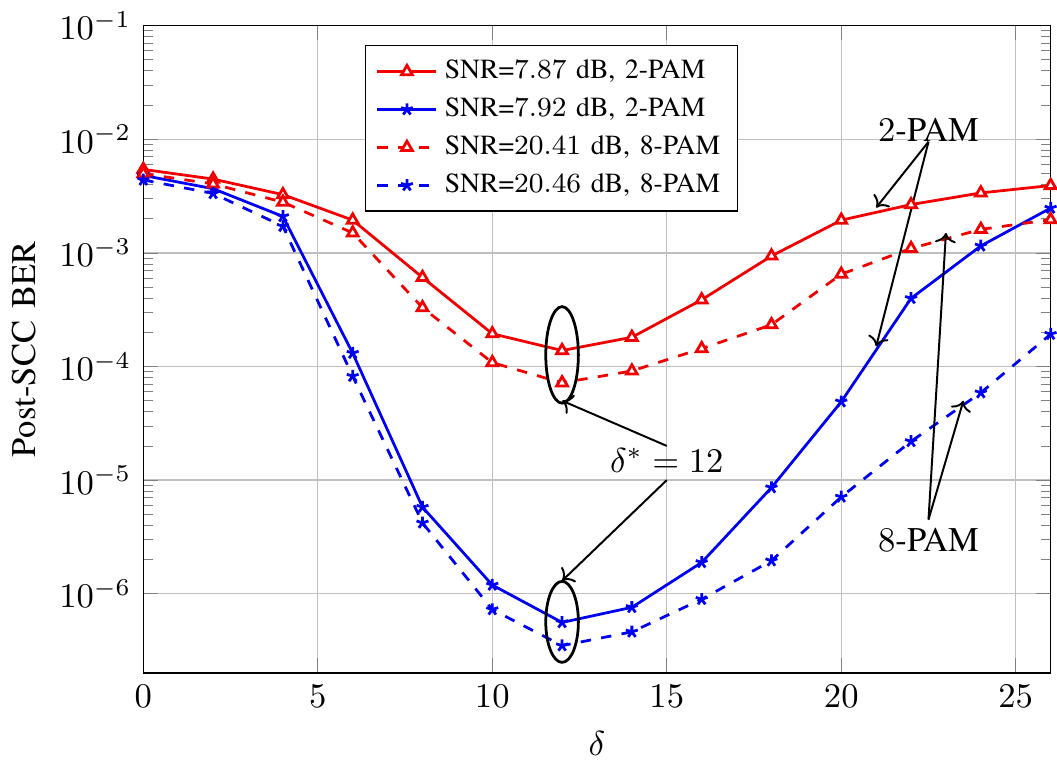}
\caption{Post-SCC BER vs. LLR threshold $\delta$ for code rate $R=0.92$. The modulation formats include $2$-PAM (solid lines) and $8$-PAM (dashed lines).}
\label{fig: Delta for R=0.92}
\end{figure}

Figs.~\ref{fig: Delta for R=0.83} and \ref{fig: Delta for R=0.92} show the post BER performance vs. $\delta$ for SCC code rates of $R=0.83$ and $0.92$, respectively. The corresponding component codes we used are BCH$(228,209,2)$ and BCH$(504,485,2)$. These parameters are obtained by shortening the extended BCH$(512,493,2)$ by $284$ and $8$ bits, respectively. We investigate the BER performance under two SNRs for each code rate. Furthermore, we investigate two modulation formats: 2-PAM (solid lines) and 8-PAM (dashed lines). The results in Figs.~\ref{fig: Delta for R=0.83} and \ref{fig: Delta for R=0.92} show that for both code rates and modulation formats, the optimum threshold is $\delta^*=12$, which is slightly larger than the one in Fig.~\ref{fig: Delta for R=0.87}. Figs.~\ref{fig: Delta for R=0.83} and \ref{fig: Delta for R=0.92} also show that SCCs with 8-PAM are less sensitive to an overestimation of the optimum value of $\delta^*$ than SCC with 2-PAM. This can observed by the relatively flat BER curves for 8-PAM when $\delta>\delta^*$.

\subsection{Post-BER Performance Analysis}

Fig. \ref{fig: SCC for m=8,t=2} shows the BER performance vs. SNR for $R=0.87$ and 2-PAM. As suggested by the results in Fig.~\ref{fig: Delta for R=0.87}, the LLR threshold to mark HRBs is set to $\delta=10$, which is the optimum value $\delta^*$ at the point of SNR$=6.98$~dB. Two baselines are: standard decoding where miscorrections are not dealt with (circles), and miscorrection-free decoding (stars). The latter is obtained via a genie BDD decoder which corrects the received sequence only when the number of errors is not more than $t$. The black dotted curve shows the estimated error floor of standard SCC decoding. It only considers the main contributor of minimal stall patterns, estimated as\cite[Sec.~V]{Smith2012}
\begin{equation}\label{eq:ErrorFloor}
    \text{BER}_{\text{post}}\approx \frac{(t+1)^2}{w^2}M_{\text{min}}\text{BER}_{\text{pre}}^{(t+1)^2}
\end{equation}
where 
\begin{equation}\nonumber
   M_{\text{min}}=\dbinom{w}{t+1}\mathlarger{\sum}_{m=1}^{t+1} {\dbinom{w}{m}\dbinom{w}{t+1-m}}.
\end{equation}
and $\text{BER}_{\text{pre}}$ is the channel error probability
\footnote{The reason why there is a quite high error floor in Fig. \ref{fig: SCC for m=8,t=2} is the realtivelt short SCC we used. Longer SCC codes (like a G.$709$-compatible one Yi Cai studied\cite{CaiYi-OFC2018,CaiYi-JLT2018}) do not have an error floor above BER$=10^{-15}$.}. This figure also shows the performance of previously proposed methods: \cite{SmithPhD} (diamonds) and \cite{Christian1} (crossed circles).

\begin{figure}[tpb]
\centering
\includegraphics[width=0.48\textwidth]{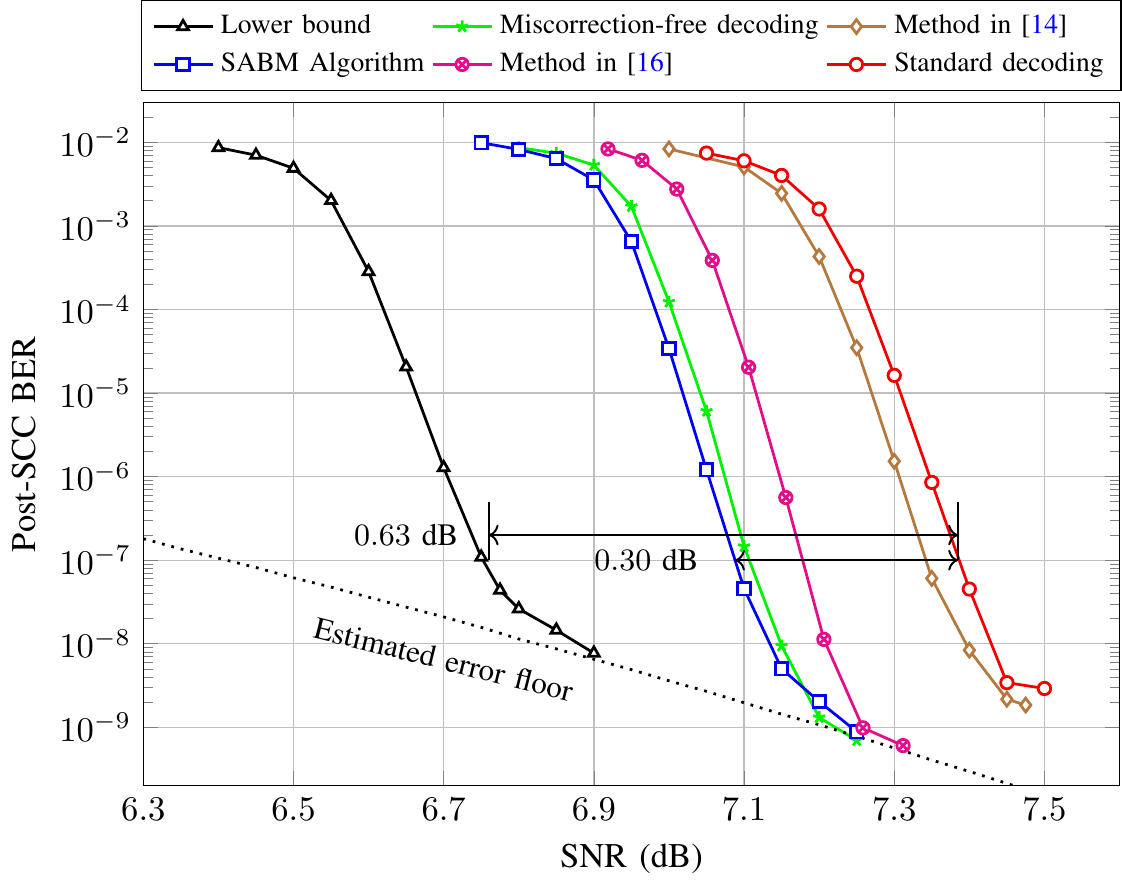}
\caption{Post-SCC BER vs. SNR for code rate $R=0.87$ and and $2$-PAM. Black dotted line is the estimated error floor of standard SCC decoding based on \eqref{eq:ErrorFloor}.}
\vspace{-2ex}
\label{fig: SCC for m=8,t=2}
\end{figure}

\begin{figure*}[tpb]
\centering
\includegraphics[width=1\textwidth]{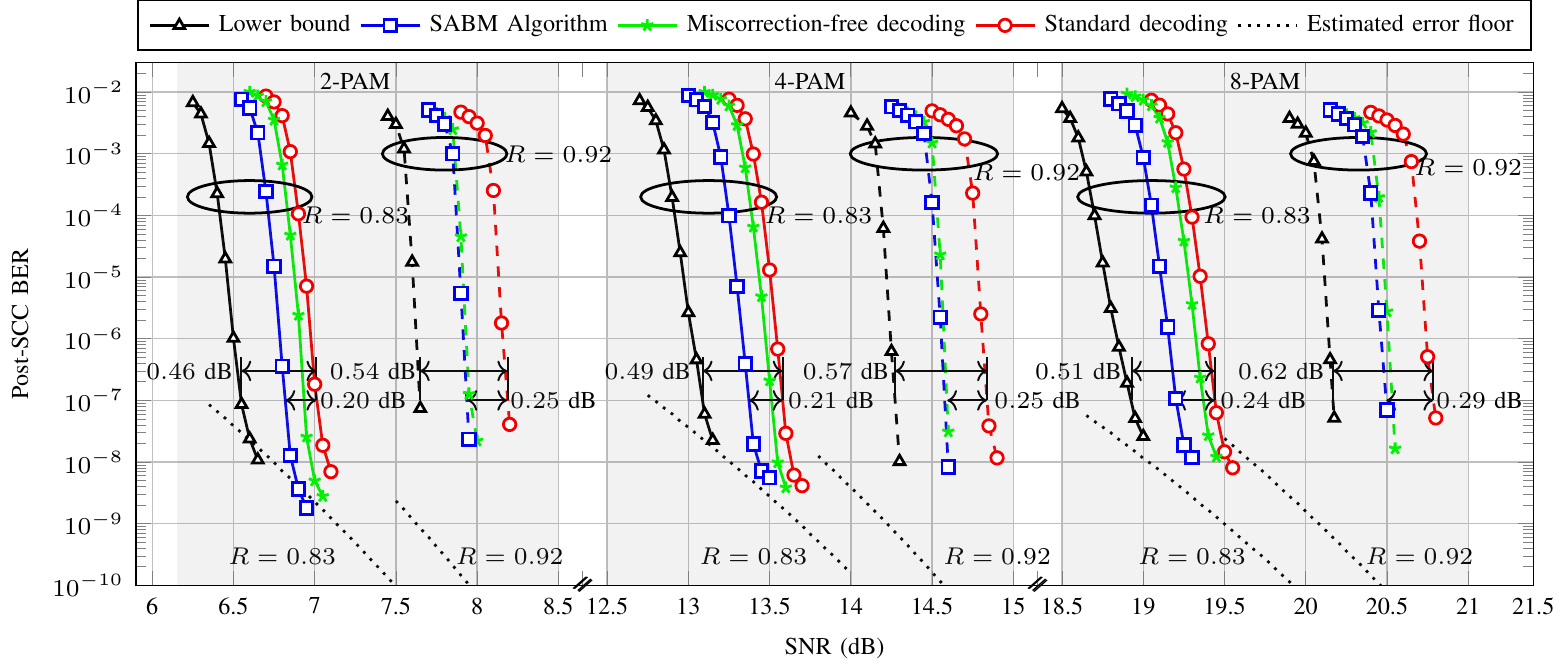}
\caption{Post-SCC BER vs. SNR for code rates $R=0.83$ (solid lines) and $R=0.92$ (dashed lines) with 2-PAM, 4-PAM, and 8-PAM modulation formats. Black dotted lines are the estimated error floors of standard SCC decoding based on \eqref{eq:ErrorFloor}.}
\label{fig: SCC for m=9,t=2}
\end{figure*}

As shown in Fig. \ref{fig: SCC for m=8,t=2}, the SABM algorithm (squares) outperforms standard decoding by $0.3$~dB and also outperforms both \cite{SmithPhD} and \cite{Christian1}. These two methods in \cite{SmithPhD} and \cite{Christian1} only prevent miscorrections, and thus, their performance is bounded by the miscorrection-free case. Although the SABM algorithm only deals with miscorrections related to the last block of each window, it outperforms the miscorrection-free case. This is due to its additional ability to better deal with miscorrections and decode even when BDD initially fails. In terms of error floor, it can be found that the performance of the SABM algorithm is lower than standard decoding, and close to the miscorrecton-free case.

Fig. \ref{fig: SCC for m=8,t=2} also shows a lower bound for the SABM algorithm (triangles). This bound is obtained by a genie decoder which emulates a best-case scenario for the SABM algorithm. This genie decoder is assumed to be able to ideally identify all miscorrections in the last two blocks of the window. This corresponds to have an idealized MD block in the top part of Fig.~\ref{fig: Method}. 
The genie decoder also emulates an idealized assumption on what the BF block in Fig.~\ref{fig: Method} can do. For this, we assume that the decoder knows exactly which bits in the last two blocks are errors. If a given sequence has $t+j$ errors ($j=1$ for Case 1, or $j=d_{0}-w_{\textrm{H}}(e)-t$ for Case 2), and at least $j$ errors are located in the last block, the genie decoder flips $j$ errors in the last block, and then the received sequence is correctly decoded. If less than $j$ errors are located in the last block, the genie decoder declares a failure. The results in Fig. \ref{fig: SCC for m=8,t=2} show that the maximum potential gain for our receiver structure (for $2$-PAM, $R=0.87$, and $t=2$) is $0.63$~dB. The SABM algorithm almost achieved half of this gain with very small added complexity (see Sec.~\ref{Sec.NumRes.Complexity} for details).

Fig.~\ref{fig: SCC for m=9,t=2} shows the simulation results of the SABM algorithm for $R=0.83$ and $0.92$. For each code rate, three modulation formats are considered: 2-PAM, 4-PAM and 8-PAM. As shown in Figs.~\ref{fig: Delta for R=0.87}--\ref{fig: Delta for R=0.92}, using an optimized $\delta^*$ for each code rate and modulation format gives the best BER. However, for simplicity, the LLR threshold we use here is set to $\delta=10$.
For SCCs with $R=0.92$, it is difficult to obtain very low BER, 
thus only the waterfall region are shown. It can be seen from Fig.~\ref{fig: SCC for m=9,t=2} that for different modulation formats and code rates, the SABM algorithm always outperforms the miscorrection-free case, also on the error floor region for $R=0.83$. When compared to standard staircase decoding, the achieved gains are between $0.20$~dB and $0.29$~dB, while the obtained maximum potential gains are between $0.46$~dB and $0.62$~dB at the BER of $10^{-7}$. The results in Fig.~\ref{fig: SCC for m=9,t=2} also show that the gains increase as the modulation size increases. 

\subsection{Complexity Analysis}\label{Sec.NumRes.Complexity}
The number of calls to the component BDD decoder is a key factor defining the complexity and latency for iterative decoding of SCCs. In order to deal with BDD decoding failures and miscorrections, the SABM algorithm needs to call the component BDD decoder multiple times (once after every BF operation). These additional calls will increase the SCC decoding complexity and latency. To quantify this, we estimate the average number of calls to the component BDD decoder within one decoding window. The relative complexity increase caused by the SABM algorithm with respect to standard SCC decoding is thus given by
\begin{align}\label{eq.eta}
\eta\triangleq\frac{\overline{N}-N_{\text{sd}}}{N_{\text{sd}}}=\frac{\overline{N}-w(L-1)\ell}{w(L-1)\ell}, 
\end{align}
where $\overline{N}$ and $N_{\text{sd}}$ are the number of BDD calls for the SABM algorithm and for the standard SCC decoding, respectively. In what follows we estimate the value of $\eta$ in \eqref{eq.eta} by estimating the average $\overline{N}$ using the first $10,000$ decoding windows.

\begin{figure}[t]
    \centering
    \captionsetup{justification = centering} 
        \label{1}
        \includegraphics[width=0.48\textwidth]{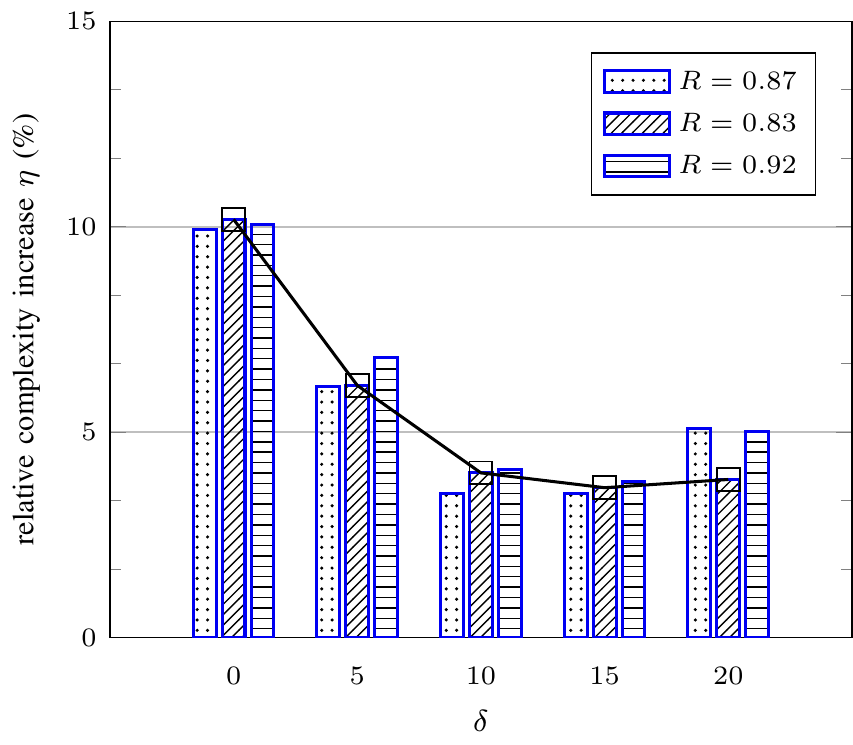}
     \caption{The relative complexity increase $\eta$ vs. $\delta$ with $L=9$, $\ell=7$ and $2$-PAM. The SNRs are $6.98$~dB, $6.72$~dB and $7.87$~dB for $R=0.87$, $0.83$ and $0.92$, respectively.}
    \label{fig:Complexity_SameSNR}
\end{figure}

Fig.~\ref{fig:Complexity_SameSNR} shows the relative complexity increase $\eta$ under different LLR threshold $\delta$. The SNRs are $6.98$~dB, $6.72$~dB and $7.87$~dB, which result in a post-SCC BER of $10^{-4}$ under $\delta=10$, 2-PAM, and code rates $R=0.87$, $0.83$ and $0.92$, respectively. The number of calls to BDD for the standard SCC decoding are $N_{\text{sd}}=7168$, $6384$ and $14112$ for code rates $R=0.87$, $0.83$ and $0.92$, respectively. The black fitted curve in Fig.~\ref{fig:Complexity_SameSNR} is used to better show the trend of the increased complexity of SCC with $R=0.83$. The other two code rates show a similar trend (not shown in this figure). The results in Fig.~\ref{fig:Complexity_SameSNR} show that the relative complexity increase around the optimum LLR threshold $\delta^*$ ($\delta^* \approx 10$ for $R=0.87$, $\delta^*=12$ for $R=0.83$ and $0.92$) is the least, and is only around $4\%$. As explained in Sec.~\ref{sec:LLRchoice}, if $\delta$ is too small, more outputs of BDD will be mistakenly identified as miscorrections. Consequently, the SABM-based staircase decoder will recall BDD for each marked miscorrection to try to decode it, thus lead to an increased additional calls to BDD. On the other hand, if $\delta$ is higher than the optimum threshold $\delta^*$, there are less bits marked as HRB, and thus, miscorrections cannot be identified effectively. More errors (caused by miscorrections) will then be added to the received sequences. As a consequence, decoding failure happens more often in the following iterations. Similarly, the SABM-based staircase decoder will recall BDD to try to decode each BDD decoding failure. Therefore, the complexity increases slightly in this case too.

\begin{figure}[t]
    \centering
    \captionsetup{justification = centering}
        \label{1}
        \includegraphics[width=0.48\textwidth]{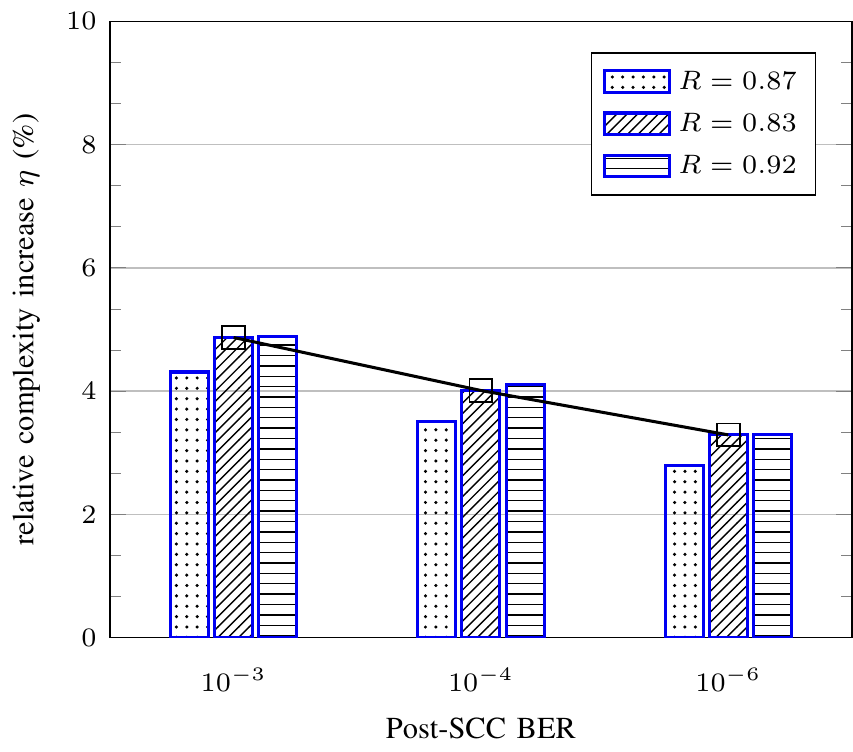}
      \caption{The relative complexity increase $\eta$ vs. post-BER with $\delta=10$, $L=9$, $\ell=7$ and $2$-PAM.}
     \label{fig:Complexity_sameDelta}
\end{figure}

Fig.~\ref{fig:Complexity_sameDelta} shows the relative complexity increase $\eta$ of the SABM algorithm under different post-SCC BER. Similarly to Fig.~\ref{fig:Complexity_SameSNR}, the black fitted curve is used to better show the trend of the increased complexity of SCC with $R=0.83$. The LLR threshold used was $\delta=10$. When the SNR increases, there are less errors in the received sequence and most of the time BDD can deal with them successfully. Therefore, the case of decoding failure or miscorrection happens less frequently, leading to a decreased additional calls to BDD in the SABM algorithm. This effect is shown in Fig.~\ref{fig:Complexity_sameDelta}, where the relative complexity increase reduces as the channel condition improves. In the asymptotic case (SNR tending to infinity), the total number of BDD calls in the SABM algorithm will approach that of the standard SCC decoding, and thus, $\eta\to 0$.

\section{Extension to Product Codes}

A product code is a set of square arrays of size $n_c\times n_c$, constructed in such a way that every row or column is an allowed codeword in some component (block) code $(n_c,k_c,t)$ \cite{Elias1954}. 
Multiple algorithms have been recently proposed to improve the decoding performance of PCs while keeping a manageable decoding complexity, as in e.g.~\cite{MittelholzerISITA2016, CondoGlobalSIP2016, SridharanPAT2001, JornTComm2011,ChristianJournal,AlirezaISTC, JornITW2007}. 
The algorithm we introduced in this paper can also be used, with slight modifications, to improve conventional decoding of PCs. In this section, first we show how to modify the SABM algorithm presented in Sec.~\ref{sec:algorithm} to suit PCs, and second we illustrate the gains achieved with this improved approach.

\subsection{Modification to the SABM Algorithm for PC Decoding}

In the SCC case, both MD and BF are applied only to the last block in the decoding window exploiting the channel reliabilities (LLRs). This is justified by the fact that the last block contains less reliable bits as no previous decoding iterations were perfomed on it. Differently from SCCs, in the PC case, row and column decoding are performed iteratively within the same block. As a result, no bits within each block can be regarded as more or less reliable than others, and conflicts between column and row decoding are likely to arise. Thus, one may expect to obtain gains only when MD and BF is performed within the first decoding iterations.

In particular, we have analyzed the performance of our algorithm and found it needs to be modified as follows. MD and BF operations should only be performed within the first decoding iteration and the first half of the second iteration (row decoding).
Extending beyond the second iteration was observed to degrade the decoding performance, hypothetically due to conflicts between row and column decodings. Furthemore, the BF is only adopted in case of decoding failure (HUB flipping) and not in the case of miscorrection. As for the row decoding operated in the first iteration, MD is only operated based on the marked HRBs, since no previous information on the codeword syndromes is available from the decoder. From the first column decoding onwards MD is based on both bit marking or syndrome information. The reliability threshold to mark the bits was also optimized for the PC case and the optimal value was found to be identical to the case of SCC, $\delta=\delta^*=10$.

\begin{figure}[t]
\centering
\includegraphics[width=0.48\textwidth]{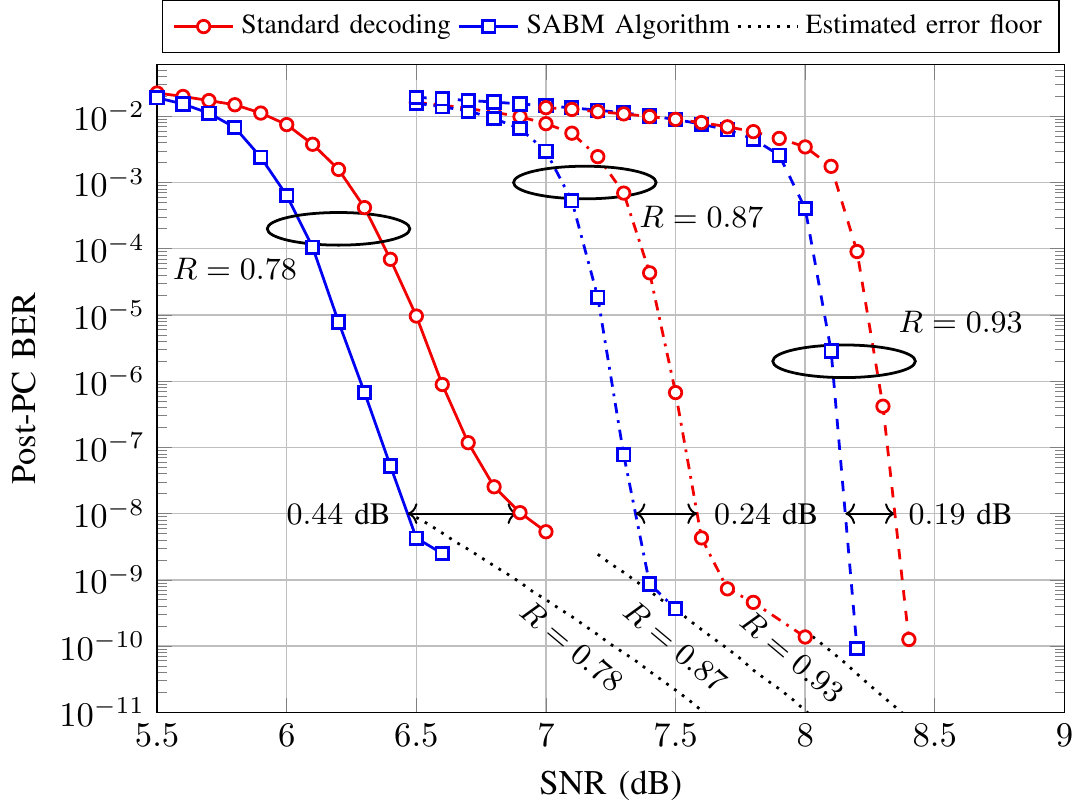}
\caption{Post-PC BER vs. SNR for code rates $R=0.78$, $0.87$ and $0.93$ and with 2-PAM. Black dotted lines are the estimated error floors of standard PC decoding.}
\label{fig:productcode}
\end{figure}

\subsection{Post-BER Performance Analysis}
We consider 3 different PCs based on $1$-bit extended BCH codes as component codes with the following parameters $(128,113,2)$, $(256,239,2)$, and $(512,493,2)$. 
These parameters result in a $128\times 128$, $256\times 256$, and $512\times 512$ PC code arrays with overall code rate $R=0.78$, $0.87$ and $0.93$, respectively. In order to compare with the algorithm proposed in \cite{ChristianJournal}, the parameters of PC with $R=0.78$ are the same as that in \cite[Fig.~2]{ChristianJournal}.

The results are shown in Fig.~\ref{fig:productcode} for an AWGN channel and for a 2-PAM modulation format. The black dotted curves show the estimated error floor calculated by using \eqref{eq:ErrorFloor} but with $M_{\text{min}}=\dbinom{w}{t+1}^2$ \cite[Eq.~7]{ChristianJournal}. For PCs with code rates of $R=0.78$, $0.87$ and $0.93$, $w=128$, $256$, and $512$, respectively. When compared to standard PC decoding, the achieved additional gains at BER of $10^{-8}$ are $0.44$~dB, $0.24$~dB, and $0.19$~dB for $R=0.78$, $0.87$ and $0.93$, respectively. In particular, the obtained $0.44$~dB additional gain of PC with $R=0.78$ is slightly larger than that (i.e., $0.40$~dB) in \cite[Fig.~2]{ChristianJournal}. What's more, the SABM algorithm is much simpler than that in \cite{ChristianJournal} as it is only applied in the first $1.5$ iterations and no need to track the change of the flipped bits.

\section{Conclusions}
In this paper, a novel decoding algorithm for staircase codes was proposed. This algorithm is based on simple modification of the standard hard-decision-based staircase decoder and relies on the idea of marking bits. The algorithm consists of an improved miscorrection-detection mechanism and a bit-flipping operation to effectively prevent miscorrections and increase the error correcting performance of bounded-distance decoding. Large gains compared to standard SCC decoding were obtained with a very low added complexity. The proposed algorithm was also extended to product codes with a similar performance improvement. Future works include a detailed implementation analaysis as well as a detailed experimental verification.

\balance

\section*{Acknowledgments}

This work is in part supported by the NSFC Program (No. 61431003, 61821001, 61625104, and 61701155). The work of A. Alvarado and G. Liga is supported by the Netherlands Organisation for Scientific Research (NWO) via the VIDI Grant ICONIC (project number 15685) and has received funding from the European Research Council (ERC) under the European Union's Horizon 2020 research and innovation programme (grant agreement No 57791). The author Yi Lei would like to thank China Scholarship Council (CSC) for supporting her study in Netherlands.

\bibliographystyle{IEEEtran}
\bibliography{IEEEabrv,refs_ISTC}

\begin{thebibliography}{10}
\providecommand{\url}[1]{#1}
\csname url@samestyle\endcsname
\providecommand{\newblock}{\relax}
\providecommand{\bibinfo}[2]{#2}
\providecommand{\BIBentrySTDinterwordspacing}{\spaceskip=0pt\relax}
\providecommand{\BIBentryALTinterwordstretchfactor}{4}
\providecommand{\BIBentryALTinterwordspacing}{\spaceskip=\fontdimen2\font plus
\BIBentryALTinterwordstretchfactor\fontdimen3\font minus
  \fontdimen4\font\relax}
\providecommand{\BIBforeignlanguage}[2]{{%
\expandafter\ifx\csname l@#1\endcsname\relax
\typeout{** WARNING: IEEEtran.bst: No hyphenation pattern has been}%
\typeout{** loaded for the language `#1'. Using the pattern for}%
\typeout{** the default language instead.}%
\else
\language=\csname l@#1\endcsname
\fi
#2}}
\providecommand{\BIBdecl}{\relax}
\BIBdecl

\bibitem{G975}
ITU, \emph{Forward error correction for submarine systems}, ITU-T
  Recommendation G.975, Oct. 2000.

\bibitem{G9751}
------, \emph{Forward error correction for high bit-rate {DWDM} submarine
  systems}, ITU-T Recommendation G.975.1, Feb. 2004.

\bibitem{400G1}
M.~Camera, B.~Olsson, and G.~Bruno, ``Beyond 100{G}bit/s: System implications
  towards 400{G} and 1{T},'' in \emph{European Conference on Optical
  Communciation (ECOC)}, Torino, Italy, Sep. 2010.

\bibitem{400G3}
X.~Zhou and L.~Nelson, ``Advanced {DSP} for 400 {G}b/s and beyond optical
  networks,'' \emph{Journal of Lightwave Technology}, vol.~32, no.~16, pp.
  2716--2725, Aug. 2014.

\bibitem{JornTComm2011}
J.~Justesen, ``Performance of product codes and related structures with
  iterated decoding,'' \emph{IEEE Transactions on Communications}, vol.~59,
  no.~2, pp. 407--415, Feb. 2011.

\bibitem{Smith2012}
B.~P. Smith, A.~Farhood, A.~Hunt, F.~R. Kschischang, and J.~Lodge, ``Staircase
  codes: {FEC} for 100 {G}b/s {OTN},'' \emph{Journal of Lightwave Technology},
  vol.~30, no.~1, pp. 110--117, Jan. 2012.

\bibitem{Zhang2014}
L.~M. Zhang and F.~R. Kschischang, ``Staircase codes with 6\% to 33\%
  overhead,'' \emph{Journal of Lightwave Technology}, vol.~32, no.~10, pp.
  1999--2002, May 2014.

\bibitem{OIF400G}
O.~I. Forum, \emph{Implementation Agreement 400{ZR}}, Optical Internetworking
  Forum, Jan. 2018.

\bibitem{G709}
ITU, \emph{OTU4 long-reach interface}, ITU-T Recommendation G.709.2/Y.1331.2,
  July 2018.

\bibitem{FougstedtJLt2019}
C.~Fougstedt and P.~Larsson-Edefors, ``Energy-efficient high-throughput {VLSI}
  architectures for product-like codes,'' \emph{Journal of Lightwave
  Technology}, vol. Early Access, 2019.

\bibitem{BarakatainJLT2018}
M.~Barakatain and F.~R. Kschischang, ``Low-complexity concatenated
  {LDPC}-staircase codes,'' \emph{Journal of Lightwave Technology}, vol.~36,
  no.~12, pp. 2443--2449, June 2018.

\bibitem{BinACP2018}
B.~Chen, Y.~Lei, D.~Lavery, C.~Okonkwo, and A.~Alvarado, ``Rate-adaptive coded
  modulation with geometrically-shaped constellations,'' in \emph{2018 Asia
  Communications and Photonics Conference (ACP)}, Hangzhou, China, Oct 2018.

\bibitem{BDD}
I.~Reed and X.~Chen, \emph{Error control coding for data network},
  1st~ed.\hskip 1em plus 0.5em minus 0.4em\relax New York: Kluwer Academic
  Publishers, 1999.

\bibitem{SmithPhD}
B.~P. Smith, ``Error-correcting codes for fibre-optic communication systems,''
  Ph.D. dissertation, University of Toronto, 2011.

\bibitem{ChristianJournal}
C.~H\"{a}ger and H.~D. Pfister, ``Approaching miscorrection-free performance of
  product codes with anchor decoding,'' \emph{IEEE Transactions on
  Communications}, vol.~66, no.~7, pp. 2797--2808, Mar. 2018.

\bibitem{Christian1}
------, ``Miscorrection-free decoding of staircase codes,'' in \emph{European
  Conference on Optical Communciation (ECOC)}, Gothenburg, Sweden, Sep. 2017.

\bibitem{Alireza}
A.~Sheikh, A.~{Graell i Amat}, and G.~Liva, ``Iterative bounded distance
  decoding of product codes with scaled reliability,'' in \emph{European
  Conference on Optical Communciation ({ECOC})}, Sep. 2018.

\bibitem{AlirezaISTC}
A.~Sheikh, A.~G.~i. Amat, G.~Liva, C.~H{\"{a}}ger, and H.~D. Pfister, ``On
  low-complexity decoding of product codes for high-throughput fiber-optic
  systems,'' in \emph{2018 IEEE 10th International Symposium on Turbo Codes \&
  Iterative Information Processing ({ISTC})}, Hong Kong, China, Dec. 2018.

\bibitem{AlirezaOFC2019}
C.~Fougstedt, A.~Sheikh, A.~{Graell i Amat}, G.~Liva, and P.~Larsson-Edefors,
  ``Energy-efficient soft-assisted product decoders,'' in \emph{Optical Fiber
  Communciations Conference and Exposition ({OFC})}, San Diego, USA, Mar. 2019.

\bibitem{AlirezaarXiv2019}
A.~{Sheikh}, A.~G.~i. {Amat}, and G.~{Liva}, ``Binary message passing decoding
  of product codes based on generalized minimum distance decoding,''
  \emph{arXiv e-prints}, Jan. 2019.

\bibitem{ChaseDecoding}
D.~Chase, ``A class of algorithms for decoding block codes with channel
  measurement information,'' \emph{IEEE Transactions on Information Theory},
  vol.~18, no.~1, pp. 170--172, Jan. 1972.

\bibitem{ErasureDecoding}
S.~Lin and D.~J. Costello, \emph{Error Control Coding}.\hskip 1em plus 0.5em
  minus 0.4em\relax New Jersey, USA: Pearson, 2004.

\bibitem{Douxin_ISTC2018}
X.~Dou, M.~Zhu, J.~Zhang, and B.~Bai, ``Soft-decision based sliding-window
  decoding of staircase codes,'' in \emph{2018 IEEE 10th International
  Symposium on Turbo Codes \& Iterative Information Processing (ISTC)}, Hong
  Kong, China, Dec. 2018.

\bibitem{YiISTC2018}
Y.~Lei, A.~Alvarado, B.~Chen, X.~Deng, Z.~Cao, J.~Li, and K.~Xu, ``Decoding
  staircase codes with marked bits,'' in \emph{2018 IEEE 10th International
  Symposium on Turbo Codes \& Iterative Information Processing ({ISTC})}, Hong
  Kong, China, Dec. 2018.

\bibitem{BinOFC2019}
B.~Chen, Y.~Lei, S.~van~der Heide, J.~van Weerdenburg, A.~Alvarado, and
  C.~Okonkwo, ``First experimental verification of improved decoding of
  staircase codes using marked bits,'' in \emph{Optical Fiber Communciations
  Conference and Exposition ({OFC})}, San Diego, USA, Mar. 2019.

\bibitem{LLR}
L.~Szczecinski and A.~Alvarado, \emph{Bit-Interleaved Coded Modulation:
  Fundamentals, Analysis, and Design}.\hskip 1em plus 0.5em minus 0.4em\relax
  Chichester, UK: Wiley, 2015.

\bibitem{CaiYi-OFC2018}
Y.~Cai, W.~Wang, W.~Qian, J.~Xing, K.~Tao, J.~Yin, S.~Zhang, M.~Lei, E.~Sun,
  K.~Y. an~Hungchang~Chien, Q.~Liao, and H.~Chen, ``{FPGA} investigation on
  error-floor performance of a concatenated staircase and hamming code for
  400{G}-{ZR} forward error correction,'' in \emph{Optical Fiber Communciations
  Conference and Exposition ({OFC})}, San Diego, USA, Mar. 2018.

\bibitem{CaiYi-JLT2018}
------, ``{FPGA} investigation on error-flare performance of a concatenated
  staircase and hamming fec code for 400{G} inter-data center interconnect,''
  \emph{Journal of Lightwave Technology}, vol. Early Access, Nov. 2018.

\bibitem{Elias1954}
P.~Elias, ``Error-free coding,'' \emph{Transactions of the IRE Professional
  Group on Information Theory}, vol.~4, no.~4, pp. 29--37, Sep. 1954.

\bibitem{MittelholzerISITA2016}
T.~Mittelholzer, T.~Parnell, N.~Papandreou, and H.~Pozidis, ``Improving the
  error-floor performance of binary half-product codes,'' in \emph{2016
  International Symposium on Information Theory and Its Applications (ISITA)},
  Oct 2016, pp. 295--299.

\bibitem{CondoGlobalSIP2016}
C.~Condo, F.~Leduc-Primeau, G.~Sarkis, P.~Giard, and W.~J. Gross, ``Stall
  pattern avoidance in polynomial product codes,'' in \emph{2016 IEEE Global
  Conference on Signal and Information Processing (GlobalSIP)}, Dec 2016, pp.
  699--702.

\bibitem{SridharanPAT2001}
\BIBentryALTinterwordspacing
M.~J. S.~Sridharan and T.~Coe, ``Product code based forward error correction
  system,'' Patent US 6,810,499 B2, 13, 2003. [Online]. Available:
  \url{http://www.patentlens.net/patentlens/...}
\BIBentrySTDinterwordspacing

\bibitem{JornITW2007}
J.~Justesen and T.~Hoholdt, ``Analysis of iterated hard decision decoding of
  product codes with reed-solomon component codes,'' in \emph{2007 IEEE
  Information Theory Workshop}, Sep. 2007, pp. 174--177.

\end{thebibliography}

\end{document}